# Revealing the Competing Contributions of Charge Carriers, Excitons, and Defects to the Non-Equilibrium Optical Properties of ZnO

*Laura Foglia[1], Sesha Vempati[1], Boubacar Tanda Bonkano[1], Lukas Gierster[1], Martin Wolf[1], Sergey Sadofev[2], and Julia Stähler[1,*]*

[1]Department of Physical Chemistry, Fritz Haber Institute of the Max Planck Society, Faradayweg 4-6, 14195 Berlin, Germany

*Corresponding author: staehler@fhi-berlin.mpg.de

[2]AG Photonik, Institut für Physik, Humboldt-Universität zu Berlin, Newtonstr. 15, 12489 Berlin, Germany

**Abstract:** Due to its wide band gap and high carrier mobility, ZnO is, among other transparent conductive oxides, an attractive material for light-harvesting and optoelectronic applications. Its functional efficiency, however, is strongly affected by defect-related in-gap states that open up extrinsic decay channels and modify relaxation timescales. As a consequence, almost every sample behaves differently, leading to irreproducible or even contradicting observations. Here, a complementary set of time-resolved spectroscopies is applied to two ZnO samples of different defect density to disentangle the competing contributions of charge carriers, excitons, and defects to the non-equilibrium dynamics after photoexcitation: Time-resolved photoluminescence, excited state transmission, and electronic sum-frequency generation. Remarkably, defects affect the transient optical properties of ZnO across more than eight orders of magnitude in time, starting with photodepletion of normally occupied defect states on femtosecond timescales, followed by the competition of free exciton emission and exciton trapping at defect sites within picoseconds, photoluminescence of defect-bound and free excitons on nanosecond timescales, and deeply trapped holes with microsecond lifetimes. These findings do not only provide the first comprehensive picture of charge and exciton relaxation pathways in ZnO, but also uncover the microscopic origin of previous conflicting observations in this challenging material and thereby offer means of overcoming its difficulties. Noteworthy, a similar competition of intrinsic and defect-related dynamics could likely also be utilized in other oxides with marked defect density as, for instance, $TiO_2$ or $SrTiO_3$.

## 1. Introduction

In the last two decades, compound semiconductors, in particular transparent conductive oxides (TCOs), were increasingly used in manifold device applications that require optical transparency as well as high carrier densities and mobilities. Nevertheless, their triumph was hampered by the persistent lack of understanding of the detailed origin of the high conductivity in these materials, partially due to the complex involvement of different types of defects.[1] Beyond their direct impact on carrier density and mobility, defects also affect the optical and opto-electronic properties of TCOs, directly reflected in the absorption and emission characteristics[2,3,4,5,6] or more subtly manifested in the lifetimes and non-equilibrium dynamics of optical excitations as, for example, in $TiO_2$[7,8] and ZnO[9].

With its 3.4 eV direct band gap, bulk exciton binding energy of 60 meV, native *n*-type doping and high conductivity, ZnO has the potential to be an excellent TCO for optoelectronic applications in the visible and UV photon energy range and for light harvesting applications.[10] Further interesting optical properties, such as birefringence and a high second order bulk non-linear optical susceptibility,[11] result from the non-centrosymmetric bulk crystal wurtzite structure (space group $C^4_{6v}$),[12] where each oxygen atom is surrounded by four zinc atoms at the corners of a tetrahedron. The functionality and efficiency of most potential ZnO-based applications rely on charge and energy transfer mechanisms at the multiple interfaces within an optoelectronic or light-harvesting device as well as on the timescales of charge carrier or exciton diffusion and relaxation processes. Exciton lifetimes and diffusion lengths as well as the free carrier conductivity can be strongly affected by the presence of deep and shallow defect-related in-gap states,[6,13] which depend on the material's growth conditions in a way that is still far from being understood.[14]

As illustrated in Figure 1c, the ZnO electronic band structure exposes a splitting of the topmost valence band (VB) into three subbands usually termed A, B and C, which is caused by a combination of crystal field and spin-orbit coupling.[15,16] Transitions between these bands and the conduction band (CB) dominate the optical absorption and reflection spectra, showing different polarization dependencies: A and B are probed using light with the electric field polarized perpendicular to the *c*-axis of ZnO at resonance energies of 3.371 and 3.378 eV, respectively, while C is observed for a field parallel to the *c*-axis and at 3.418 eV.[17]

Additionally, most ZnO samples exhibit a green emission band attributed to both oxygen and zinc vacancies,[13,18,19,20,21] while oxygen excess results in an emission band centered at 1.8 eV.[22] Oxygen vacancies and zinc interstitials were historically also suspected to be the cause of the *n*-type conductivity in ZnO.[14] However, consensus seems to emerge that the *n*-type conductivity is mainly caused by interstitial or substitutional hydrogen.[23] When adsorbed on the surfaces, hydrogen is also responsible of downward surface band bending and the formation of a charge accumulation layer, leading to surface metallicity.[24,25,26]

Due to the native *n*-type doping, the defects lie below the Fermi energy $E_F$ in equilibrium, i.e. the defect states are normally occupied. Absorption occurs to the CB of ZnO, and emission in the visible is a result of the recombination of defect excitons $DX_{VIS}$ (electrons bound to holes located at defect sites). Beyond these effects on the optical properties of ZnO in the visible spectral range, defects also act as traps for free excitons $FX_{UV}$. The resulting defect-bound excitons $DX_{UV}$ emit light in the UV upon recombination, and the character of the defect (ionized/neutral, donor/acceptor etc.) determines the exact exciton resonance energy.[6] A related effect was recently found for the hydrogen-terminated ZnO(10-10) surface, were surface excitons localize within only 200 fs next to potential minima caused by the positively charged hydrogen adatoms.[51]

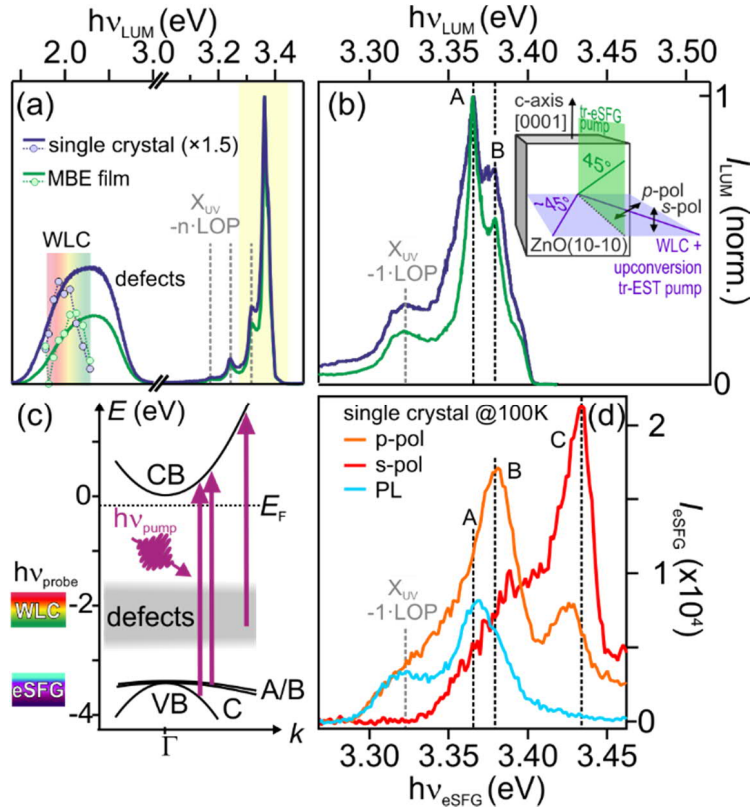

**Figure 1.** (a) PL spectra of the ZnO single crystal (blue) and MBE-grown film (green) in the visible and near-UV spectral range, normalized to the maximum emission. Markers represent photoinduced transient transmission signals in the visible range and probed by a white light continuum (WLC), discussed in section 2.2. (b) High-resolution spectra of the UV emission region and sketch of the experimental geometry. (c) ZnO electronic band structure scheme, excitation transitions, and energy regions accessed by WLC and eSFG photons through transitions near the CB minimum (left). (d) eSFG spectra of the ZnO single crystal for the two polarization configurations (*p*- in orange and *s*-polarization in red) compared to the PL spectrum (light blue).

As defects affect the optical and electronic properties of ZnO so significantly, caution is required when assigning and interpreting spectroscopic results. This is particularly the case when considering optical data where the populations are deduced from changes in the intensity and energy position of optical resonances. Indeed, within the vast amount of literature on static and non-equilibrium optical properties of ZnO, the assignment of the reported values to physical phenomena is often conflicting.[27,28,29,30,31,32,33,34,35,36] The discrepancies are likely resulting from the application of different experimental techniques to likewise different samples. For example, the literature values for linear and non-linear optical susceptibilities,[28,29] Mott density and Bohr exciton radius can diverge by as much as two orders of magnitude.[33 and references therein] Also, the reported timescales for exciton and charge carrier trapping by defects range from hundreds of fs to ns,[6,30,31,32] holes are trapped at oxygen defects faster than 80 ps,[9] the exciton lifetime is measured between tens and thousands of ps,[6,32,34,35,36] while THz spectroscopy determines the formation of a bound exciton population to occur within 200 ps.[34]

In this article, we address these issues by investigating the charge carrier and exciton dynamics in bulk ZnO using a complementary set of optical techniques that we apply to two different samples with markedly different defect density: a comparably defect-rich commercial single crystal and a high-quality thin film grown by molecular beam epitaxy (MBE). In particular, we study the time-dependent light emission of these samples after above band gap photoexcitation using time-resolved photoluminescence spectroscopy (tr-PL) based on time-correlated single photon counting (TCSPC), investigate the ultrafast charge carrier dynamics by time-resolved excited state transmission (tr-EST) experiments[37], and monitor the exciton ground state formation dynamics by time-resolved electronic sum-frequency generation (tr-eSFG). This complementary approach, combined with the comparison of the dynamics in the two different samples, provides a comprehensive picture of all relevant relaxation processes following photoexcitation. The static characterization of the optical properties of both samples reveals a defect band from 1.5 to 2.8 eV below the CB minimum and shows the three exciton-polariton resonances around 3.4 eV which were introduced above. Photoexcitation with photon energies significantly above the band gap energy (i) partially depletes defect states on femtosecond (fs) timescales and (ii) excites quasi-free carriers to the CB and VB, which decay within few to hundreds of picoseconds (ps). (iii) Quasi-free electrons bind to the ionized defects to form $DX_{VIS}$ that emit visible light on nanosecond (ns) and microsecond (µs) timescales. (iv) Finally, $FX_{UV}$ formation and decay dynamics compete with exciton trapping at defect sites on sub- or few-ns timescales. Therefore, the defect density determines whether the UV light emission occurs quickly through free exciton emission or whether defect-trapped excitons $DX_{UV}$ are formed with significantly longer recombination times.

In summary, our study shows that defects affect the quasiparticle dynamics and optical properties of ZnO in a complex way and across more than eight orders of magnitude in time, from fs to µs. It should be noted that a detailed characterization of the involved defects, their absolute densities, and environments are out of the scope of this article. Instead, this work provides a novel understanding of *when* and *how* defects interact with photoexcited carriers and excitons by showing that, depending on the defect density, free carriers are captured at defect sites to form defect(-bound) excitons *before* free excitons are generated. This unveils the microscopic origin of the previous conflicting observations and, moreover, offers opportunities to reduce or even make use of the impact of defects on the optical properties of ZnO, depending on the envisioned functionality. Lastly, our results may also contribute to a better understanding of quasiparticle dynamics in compound semiconductors in general, as intrinsic and defect-related processes compete in any material with relevant defect densities.

## 2. Results

### 2.1 Spectral properties of ZnO

As a first step, we identify the important spectral signatures of ZnO. Figure 1c summarizes the relevant electronic structure of ZnO and illustrates the possible excitation with pump photons ($h\nu_{pump}$) with energies above the ZnO band gap. Two types of vertical transitions are possible: (i) Photodepletion of normally occupied defect states forming $D^+$ and (ii) excitation of electrons in VB

A-C to the CB. Figure 1a depicts PL spectra of the single crystal sample (blue) and the MBE film (green). The two spectra are normalized at maximum intensity. Both samples exhibit a well-defined emission at 3.36 eV and clear vibronic replicas with an average separation of 66 ± 15 meV. The main peak is attributed to the radiative recombination of the bulk $FX_{UV}$ and defect-bound excitons $DX_{UV}$ and the replica separation is in good agreement with exciton scattering by longitudinal optical phonons (LOP) with the previously determined energy of 72 meV.[38,39,40,41] Additionally, both samples show a broad emission band in the visible, from 1.5 to 2.8 eV. This band is associated with the radiative recombination of defect-bound excitons $DX_{VIS}$ where electrons recombine with holes in the defect band. The relative intensity of the $DX_{VIS}$ emission from the single crystal is twice as high as in the MBE film, and the different shapes of the defect bands are likely caused by different distributions of deeply bound defects in the two samples.

Figure 1b shows a high resolution spectrum of the UV emission band, in the energy region indicated by the yellow rectangle in Figure 1a. The main emission line exhibits a substructure: Two peaks, labeled A and B, are observed at 3.365(3) and 3.379(5) eV, respectively, in good agreement with previous observations of the A and B exciton emission.[42] The UV emission linewidth is much broader in the single crystal, which suggests a larger number of defect-bound excitons $DX_{UV}$ than in the MBE film.

For the following time-resolved experiments, we used a femtosecond laser system providing ultrashort laser pulses with a spectral band width on the order of 70 meV. This band width further broadens our spectra, as shown in Figure 1d, where we compare a luminescence spectrum (light blue) of the single crystal sample to eSFG spectra measured for two polarization configurations as illustrated by the inset (red spectrum: *s*, orange spectrum: *p*). In eSFG,[43] a white light continuum in the visible range (1.8 - 2.3 eV, cf. Figure 1c) is up-converted with 1.55 eV (800 nm) photons, covering an eSFG photon energy range from 3.35 to 3.85 eV. The sum frequency intensity is enhanced whenever either the single photon energies or their sum match an absorption resonance. The polarization configurations indicate the polarization of the up-converting beam, while the white light is always *s*-polarized, as discussed in detail in the experimental section in Appendix A. At photon energies below 3.5 eV and for the *p*-polarization configuration, the eSFG shows similar spectral signatures as the PL. Note that the intensity ratio of A and B is reversed with respect to the luminescence spectrum, suggesting different matrix elements for optical transitions in the two techniques. In addition, another eSFG resonance (C) is observed at 3.43 eV. In *s*-polarization configuration, i.e. when both electric fields are parallel to the *c*-axis, instead, no A and B resonances are observed, but a significantly enhanced resonance C. Considering the good agreement of both energy position and polarization dependence of the eSFG spectra with the previously published absorption resonances of the split VBs discussed above,[17] we assign the low-energy resonances to transitions from VB A and B and the high-energy one to transitions from VB C.

## 2.2 Defect-related exciton dynamics from pico- to microseconds

The radiative recombination lifetime of the observed transitions can be measured in a TCSPC experiment, where the emission time of a photon is related to the arrival time of the pump pulse at the sample. Figure 2a shows two TCSPC traces (raw data) acquired at 3.35±0.05 eV (blue: single crystal, green: MBE film). Due to limited spectral resolution (FWHM = 100 meV), these dynamics include the emission of both, the free A and B (A/B) excitons, as well as the emission arising from the recombination of defect-bound excitons $DX_{UV}$. All TCSPC datasets can be fitted using a sum of exponential decay functions (red curves) and the best-fit parameters are listed in Appendix B. It is noteworthy that – due the large impact of defects on the dynamics – the specific time constants only have a very limited meaning. For this reason, we focus on the general trends and relative changes as well as on orders of magnitude in the following discussion. For both samples we observe PL decay times exceeding 10 ns, however in the case of the single crystal these long-lived dynamics have a more than one order of magnitude higher relative intensity (Figure 2a, right) compared to the fast contributions ($\tau < 10$ ns). In addition, the early time dynamics differ substantially for the two samples, with a large amplitude, few ns decay in the case of the MBE film dominating the dynamics.

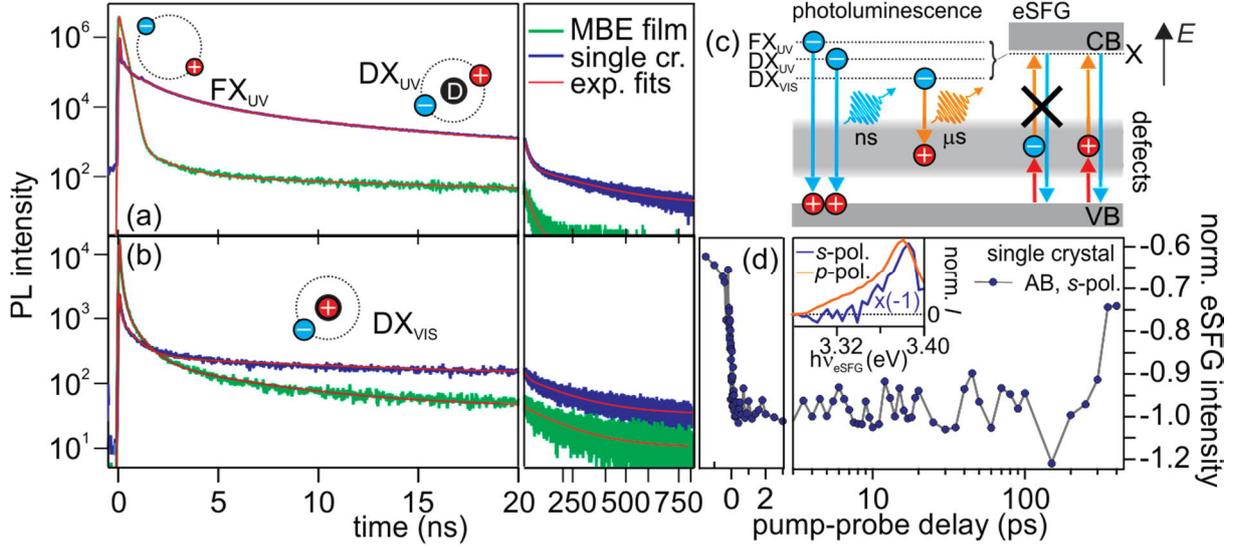

**Figure 2** (a) TCSPC at 3.35(5) eV emission energy for the MBE film (green) and the single crystal (blue) with multi-exponential fits (red, see Appendix B for details); cartoons illustrate the responsible excitons recombining faster than 50 ns. (b) TSCPC at 2.25 eV emission energy and multi-exponential fits showing radiative relaxation dynamics of defect-bound excitons extending to the μs range. (c) Scheme of the involved optical transitions. (d) Integrated eSFG intensity at the A/B-resonance energy in *s*-polarization configuration. Inset: Pump-induced eSFG of the A/B-resonance for $t < 0$ in *s*-polarization configuration compared to unpumped signal probed in *p*-polarization configuration.

We compare the dynamics of the UV emission to the radiative recombination of electron-hole pairs in the visible range (Figure 2b), acquired at an exemplary defect exciton emission energy of 2.25 eV. It should be noted that the observed dynamics are qualitatively independent of the photon energy of the emitted light and of the fluence used (0.12 – 1.8 and 0.23 – 2.0 mJ/cm$^2$ for the single crystal and MBE grown sample, respectively, not shown). The observed time evolution of this defect emission is very similar for both samples, albeit with differing relative amplitudes of the early ($\tau < 10$ ns) and late dynamics ($\tau > 10$ ns). In the case of the single crystal, the late time dynamics have a more than one order of magnitude larger relative intensity compared to the fast dynamics than in the MBE film. In both cases, the initial PL decay occurs on a sub and few ns timescale and is followed by emission extending further than the 800 ns time window accessed by our experiment (Figure 2b, right). We conclude that the visible light emission is resulting from the recombination of defect-bound excitons $DX_{VIS}$ where the hole is located in the defect band (cf. Figure 2c). The emission occurs on multiple, long timescales. Combined with the spectral broadness of the emission lines this suggests an inhomogeneous distribution of normally occupied defect sites in the ZnO, possibly due to varying defect environments. The higher relative intensity of the long-lived defect emission in the case of the single crystal is consistent with a larger number of deeply bound $DX_{VIS}$ in the single crystal.

Based on these observations and considering that Figure 2a shows emission dynamics of free (A/B) *and* defect-bound excitons, we interpret the UV emission dynamics as follows: As the amplitude of the slowly decaying UV PL in the defect-rich single crystal sample exhibits a more than 10 times larger relative intensity, it seems highly likely that this signal is resulting from radiative recombination of formally free excitons that are trapped near a localized defect site as illustrated by the right cartoon in Figure 2a. Such defect-bound excitons $DX_{UV}$ have a higher binding energy than the $FX_{UV}$ (cf. Figure 2c)[6,42]. In this scenario, only the fast emission dynamics on few- or sub-ns timescales results from $FX_{UV}$ emission and is limited by exciton trapping at defects.

Complementary to the PL experiments described above, we address the defect dynamics also in a tr-eSFG experiment.[44] In time-resolved optical spectroscopy, transient changes to the optical properties of ZnO are monitored in a so-called "pump-probe" scheme. The system is driven out of equilibrium by a first intense laser pulse (pump) that excites the sample and creates non-equilibrium conditions. The time-dependent changes of the optical properties are detected by a second laser pulse or pulse pair (probe), which arrives at the sample at a variable pump-probe time delay $\Delta t$. In the case of tr-eSFG, the probe consists of a white light continuum and a 1.55 eV gate pulse that are overlapped in time as introduced in section 2.1.

Defect dynamics can be addressed in eSFG when using the *s*-polarization configuration, i.e. when white light and the up-converting beam are both *s*-polarized and, therefore, parallel to the *c*-axis of the sample. For this configuration, no A/B resonance occurs in static eSFG spectroscopy (cf. Figure 1d) as discussed in section 2.1. As illustrated in Figure 2c, resonant eSFG enhancement in this polarization configuration does not occur when the real intermediate states in the defect band are occupied. By photoexcitation of the sample, however, we change the occupation of the involved energy levels and, therefore, the formerly forbidden transition can become allowed.

Monitoring the eSFG intensity in *s*-polarization configuration (see Figure 2d) around the A/B-resonance (3.35 - 3.4 eV) yields a distinct step function centered at $\Delta t = 0$ when pump and probe laser pulses overlap. Remarkably, even at negative pump-probe delay times, i.e. when the probe pulse arrives 25 µs (inverse of the repetition rate of the laser system at 40 kHz) after the pump, a finite negative eSFG intensity is observed, which is strengthened for $\Delta t > 0$. This means that the photoinduced eSFG intensity at the A/B resonance is very long-lived. The inset presents the pump-induced eSFG intensity distribution at negative delays (blue) and compares it to the signature of the A/B resonance as probed by static eSFG in *p*-polarization configuration (orange). Clearly, the two traces are in very good agreement.[45] Based on this, we conclude that photoexcitation of ZnO modifies either initial, intermediate, or final state of the eSFG process such that sum frequency generation at the A/B resonance becomes allowed in s-polarization configuration. As neither depleted initial nor occupied final states of the A/B exciton-polariton are expected to exhibit lifetimes exceeding 25 µs, we interpret the extremely long-lived eSFG signal as resulting from depleted (intermediate) defect states. These enable the detection of the A/B-resonance in *s*-polarized eSFG as illustrated in Figure 2c and demonstrate that positively charged defects can exhibit lifetimes larger than 25 µs.

## 2.3 Ultrafast charge carrier dynamics

Complementary to the exciton and defect dynamics described above, we now address the dynamics of charge carriers after photoexcitation and on sub-ns timescales by time-resolved transmission spectroscopy. In these experiments, the samples were photoexcited using a photon energy of 3.8 eV and with an excitation density of $8.6 \times 10^{18}$ cm$^{-3}$, i.e. close to the Mott density in ZnO, and the transient transmission changes were probed by a white light continuum, such that the probe energy was lower than the intrinsic optical band gap of ZnO. The energy level diagram in Figure 3 shows possible resonant transitions in ZnO that are accessible to the white light continuum. A *reduction* of the transmitted intensity compared to the equilibrium transmission is expected when the pump pulse induces absorption by the generation of holes in the (i) valence and (ii) defect band. While hole lifetimes in the VB are typically in the picosecond time range, the previous section showed that holes in the defect band exhibit much longer lifetimes from nanoseconds to microseconds. We emphasize that the ZnO conduction bands do not offer optical transitions in the visible range that would lead to induced absorption due to carriers at the conduction band minimum to higher lying states.[46]

On the contrary, *enhanced* transmission can either be caused by (iii) initial state bleaching, i.e. the depopulation of states in the normally occupied defect band, which inhibits absorption from these in-gap states to the CB, or by (iv) final state blocking where the electron population in the CB prevents transitions from occupied defect states. While the timescale of the former process is determined by the ns to µs hole lifetime in the defect band, the dynamics of the latter transitions are dominated by the lifetimes of electrons in the CB, which, usually, are on the ps timescale.

Beyond the effect of optical resonances, the transient transmission signal can also be influenced by photoinduced changes to the Drude response of a material.[47] While absent in the semiconducting ground state, sufficiently strongly photoexcited semiconductors can exhibit metal-like properties due to the large number of quasi-free electrons and holes in the CB and VB, respectively. The increase of free charges induces a broadband reflectivity decrease[47,48] that recovers as soon as the charges are no longer free. In addition to the above-described blocking of optical transitions ((iii) and (iv)), a transient transmission increase can, therefore, also be associated to a decrease in reflectivity if absorption does not simultaneously increase.

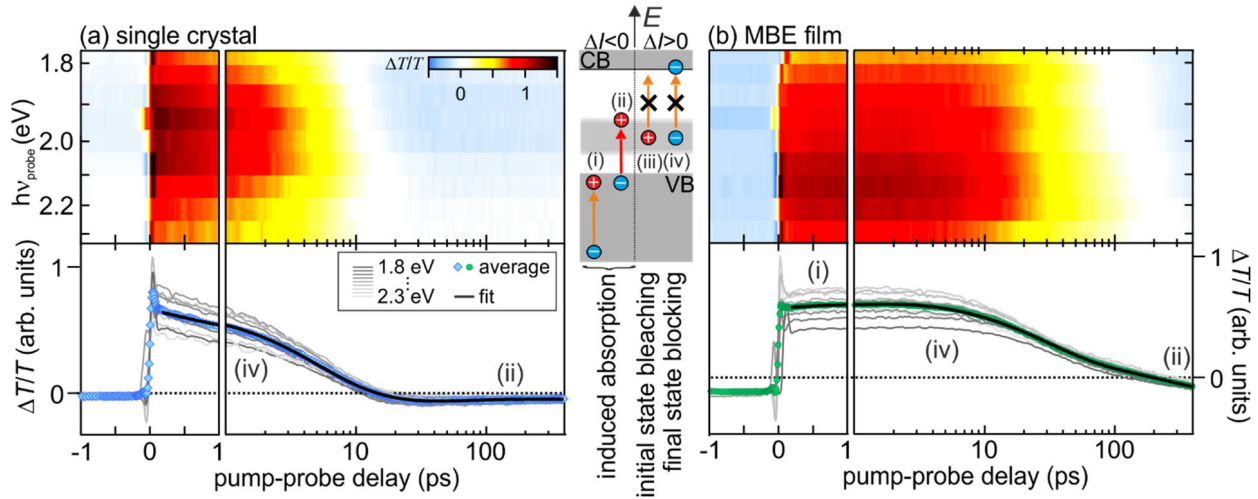

**Figure 3.** tr-EST intensity in false colors, as a function of photon energy and pump probe delay, measured using (a) the single crystal and (b) the MBE film. Bottom panels depict lineouts (gray) and average time-dependencies (markers) combined with multi-exponential fits to the data (black). The energy level diagram illustrates possible resonances responsible for decreased/increased transmission (left/right). The tr-EST is dominated by a photoinduced transparency decay with defect density-dependent ps time constants that results from the relaxation of free CB electrons.

Figure 3a and 3b depict the transient evolution of the excited state transmission (EST) of the single crystal and the MBE film in false colors, as a function of probe photon energy and pump-probe delay, respectively. Note the logarithmic time axis in the 1 to 400 ps range. The samples show an instantaneous broadband transmission *increase* upon photoexcitation and relaxation dynamics on picosecond timescales, which are faster in the case of the single crystal sample. As discussed above, this induced transparency may result from the Drude response of free carriers or from photoinduced blocking of optical resonances due to initial state bleaching (iii) or final state blocking (iv).

The bottom panels in Figure 3 display the transient EST intensity evolution for all photon energies used in the experiment (gray). Clearly, besides varying amplitudes, the time constants of the respective samples do not differ significantly. For a quantitative comparison of the dynamics, we average over all time traces (blue and green markers) and fit multiexponential decay functions to the data. In order to exclude coherent effects close to time zero when pump and probe laser pulses overlap, the fits (black) start at a pump-probe time delay of 180 fs. Note that the fits cover more than three orders of magnitude in time.

The ultrafast dynamics in both samples are dominated by a positive amplitude signal that decays double exponentially with time constants of 4.3(5) and 10.0(4) ps for the single crystal and 27(3) and 260(10) ps for the MBE film, respectively. Clearly, these time constants are significantly smaller than the ns to µs dynamics of the hole population in the defect band (cf. section 2.2). This excludes initial state bleaching as the cause of the induced transparency. On the contrary, these time constants are in good agreement with induced transparency due to final state blocking of the CB minimum (iv) through quasi-free electrons as well as the dynamics expected for the Drude response of free carriers. In both cases, the origin of the ps dynamics lies in the decay of quasi-free carriers in the CB. Whether one or the other process contributes mainly to the induced transparency can be estimated by comparing the spectral distribution of the induced transparency at $t = 100$ fs (markers) to the spectral shape of the defect band as probed by photoluminescence, which is shown in Figure 1a. Remarkably, the two complementary experiments show coinciding intensity distributions below $h\nu_{probe}= 2.1$ eV. This observation strongly suggests that the positive tr-EST signal is dominantly caused by final state (Pauli) blocking of optical transitions from the defect band to the CB as theoretically discussed for highly chemically doped ZnO,[49] rather than the Drude response of free carriers.[50] Importantly, in both cases, the picosecond dynamics of the induced transparency can be interpreted as average characteristic times of $DX_{VIS}$ and $DX_{UV}$ electron-hole pair formation. The observation of a much faster induced transparency decay in the single crystal is consistent with the higher defect density and, thus, trapping probability compared to the MBE sample and is in agreement with the PL line shapes observed in Figure 1b and the relative intensities of the ns-recombination dynamics shown in Figure 2. We conclude that the picosecond lifetime of quasi-free electrons in the ZnO CB is dominated by trapping at defect sites which is – depending on the defect density – compatible with $FX_{UV}$ ground state formation as discussed later in section 2.4. Possibly, this trapping at defect sites is a similar mechanism as the previously observed surface exciton localization near hydrogen adsorption sites.[51]

Beyond the induced transparency, both samples also exhibit reduced tr-EST signals on different timescales, albeit with significantly lower intensities. For both samples, a long-lived negative intensity is observed that exceeds the inverse repetition rate of the laser system of 5 µs, similar to the defect-induced observation of the A/B-resonance in eSFG discussed in the previous section. This signal is strongest for the lowest probe photon energies, as apparent from the light blue intensities in the false color plots in Figure 3. We interpret this signal as (ii) induced absorption from the VB into photodepleted holes in the defect band. As the defect band extends from 1.5 to 2.8 eV below the CB minimum, only the red part of our white light continuum can probe transitions from the VB maximum (3.4 eV), which exhibits the highest DOS, to defect states as illustrated by the energy level diagram in Figure 3.

The MBE sample additionally exposes photoinduced reduced transparency that decays with a time constant of 1.3 ps and is reflected as a delayed rise of the positive EST intensity in the bottom panel of Figure 3b. Possibly, this very weak signal is a result of induced absorption by transitions from lower-lying bands to photogenerated holes at the top (A/B) VBs. The higher defect density in the single crystal sample compared to the MBE film leads to a smaller fraction of photoholes in the VB at similar excitation densities, which could explain the absence of induced absorption to the top VB in tr-EST of the single crystal.

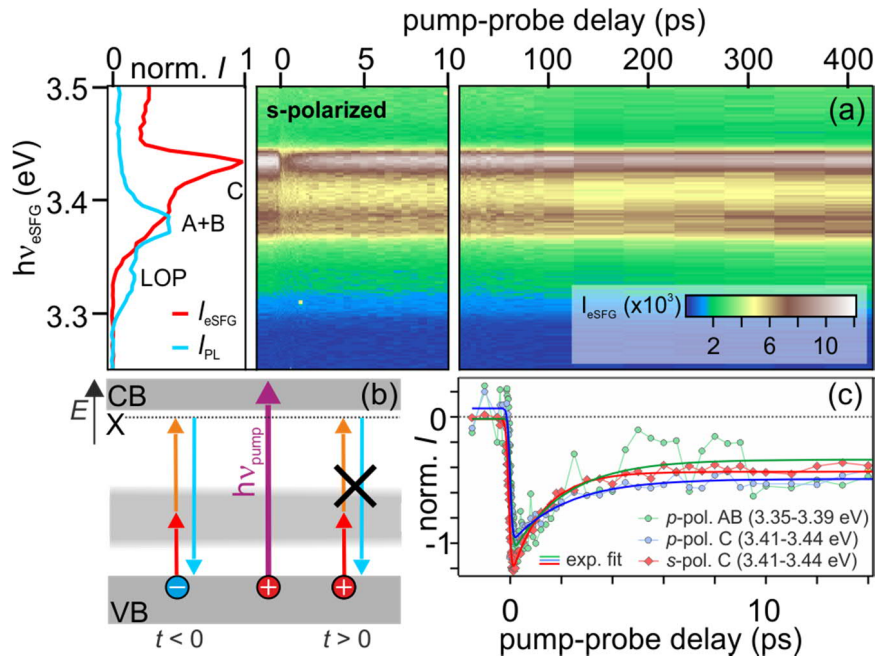

**Figure 4** (a) tr-eSFG intensity evolution of the ZnO single crystal in *s*-polarized configuration as a function of pump-probe time delay and eSFG photon energy after photoexcitation with $h\nu_{pump}$ = 4.1 eV. Left: Static eSFG (red) and luminescence (light blue) background. Reproduced from [L. Foglia, M. Wolf, J. Stähler, *Appl. Phys. Lett.* **2016**, *109*, 202106], with the permission of AIP Publishing. (b) Energy level scheme illustrating the different initial state population in the eSFG process before and after photoexcitation. (c) eSFG intensity evolution of the exciton-polariton resonance C in *s*- (red diamonds) and *p*-polarization configuration (blue circles) compared to the dynamics of the A/B resonance (green circles) and best fits (solid curves). All traces show comparable hole relaxation times.

As the tr-EST dynamics are clearly dominated by transitions from the defect band to the CB minimum, we use tr-eSFG to characterize the hole relaxation dynamics at the top of the VB as discussed in the following. Figure 4a shows the time-dependent eSFG intensity in *s*-polarization configuration (raw data) in false colors and as a function of eSFG photon energy (left axis). As in the static experiment (left panel, red curve), the time-dependent data is dominated by the C exciton resonance. Photoexcitation with $h\nu_{pump}$ = 4.1 eV leads to an abrupt depletion of the C resonance that shows an initial recovery on picosecond timescales. Integration of the eSFG intensity between 3.41 and 3.44 eV yields the intensity evolution shown in Figure 4c (red diamonds). Analogous dynamics are observed for the A/B and C resonances measured in *p*-polarization configuration (green and blue circles, Figure 4c). Clearly, all traces show coinciding relative amplitudes and comparable dynamics followed by a long-lived relaxation of the eSFG intensity. The data can be fitted (solid curves in Figure 4) using

(1) $$I(t) = \left(\text{erf}\left(\frac{t}{\tau_0}\right) + 1\right) \cdot \left(A \cdot \exp\left(-\frac{t}{\tau}\right) + y_0\right)$$

with a time-resolution limited signal depletion described by the error function, an exponential recovery with the time constant $\tau$, and a constant offset $y_0$ that accounts for the long-lived dynamics. This analysis yields consistent few ps decay constants of 2.1(7) ps, 2.2(8) ps, and 1.27(7) ps for the dynamics of the AB and C resonance (in *p*- and *s*-polarization configuration), respectively.

In contrast to the C resonance, the A/B resonance energy clearly lies below the free particle band gap of ZnO and, thus, transitions from the A/B valence bands at the resonance energy of 3.37 eV cannot involve the (free particle) CBM. Assuming a common source for the coinciding response at the A/B and C resonance energy, we conclude that this source cannot be electron dynamics in the CBM.

As the eSFG resonances show an <u>abrupt</u> depletion upon photoexcitation, the common source of the subsequent ps dynamics must originate from an excitation that is <u>immediately</u> caused by the pump pulse. Since none of the A-C excitons can be abruptly formed by the above-resonance photoexcitation at 4.1 eV, final state blocking can be omitted as the cause of the observed dynamics. Photoinduced depopulation of intermediate states in the defect band on the other hand indeed occurs abruptly at $t = 0$. However, it can neither be the origin of the ps dynamics, as depleted defects show relaxation dynamics on ns and μs timescales (cf. section 2.2). As the ps-dynamics of the resonances A-C do not result from dynamics in the intermediate nor final state of the eSFG process, we conclude that initial state bleaching through photoholes in the VBs A-C lies at the bottom of the observed dynamics.

The above interpretation is not in conflict with the 4.3 ps quasi-free electron decay observed by tr-EST of the single crystal sample, which is dominated by electron trapping in defect(-bound) excitons. However, interestingly, the few ps relaxation time of the hole population is in good agreement with the induced absorption observed in the tr-EST experiment using the MBE film discussed further above. Although the two different experiments were performed using different excitation densities, this agreement of time constants for the single crystal and the MBE-grown sample could suggest that the picosecond hole relaxation dynamics result from intrinsic relaxation processes (e.g. electron-phonon scattering or hole polaron formation) that are unaffected by the defect density of the samples. Further, it is noteworthy that the dynamics observed in the exciton-polariton resonances A/B and C do not provide comprehensive insight into all relaxation dynamics of holes at the top of the VB, but are solely sensitive to photoholes that influence the eSFG probability at the respective resonance.

**2.4 Exciton ground state formation dynamics**

As discussed above, the eSFG intensity is affected by changes to the population of initial, intermediate, and final states that alter the transition probability in the up-conversion process. The intensity evolution of the three exciton-polariton resonances that are directly probed in static eSFG (A/B in *p*-polarization configuration and C for both polarization settings) should, thus, be influenced by the formation of the exciton (1s) ground state. Indeed, in addition to the ultrafast hole relaxation dynamics discussed in the previous section, all directly probed resonances exhibit a *decrease* of the eSFG intensity on long ps timescales. This is evidenced in Figure 5, which compares the pump-induced eSFG spectra of the A/B resonance at pump-probe time delays of 20 ps, 70 ps, and 400 ps in panel (a) to the PL induced by the pump pulse (light blue) and the static eSFG background (orange) in panel (b). These spectra demonstrate that the eSFG intensity is reduced on a 100 ps timescale after off-resonant photoexcitation of ZnO. It seems highly likely that this delayed decrease of the exciton resonance intensity is associated with the formation of a ground state exciton population where, as in a common transient absorption experiment, the progressive filling of the final state leads to a reduced transition probability.

We test this hypothesis by inspecting the luminescence peak that results from excitons that couple to the longitudinal optical phonon of ZnO around 70 meV, which were introduced in section 2.1. This LOP replica of the main emission lines is absent in the static eSFG spectrum in *s*-polarization configuration (cf. Figure 1d), as this would require coupling to the LOP in the eSFG process. Time-resolved PL experiments of a photoexcited sample require special schemes such as time-correlated single photon counting or the gating with another synchronized pulse, neither of which is present in the tr-eSFG experiment. Therefore, since luminescence is a spontaneous process, it can usually not be time-resolved in an optical pump-probe experiment, but it appears as a constant background, unless stimulated emission occurs as discussed below.

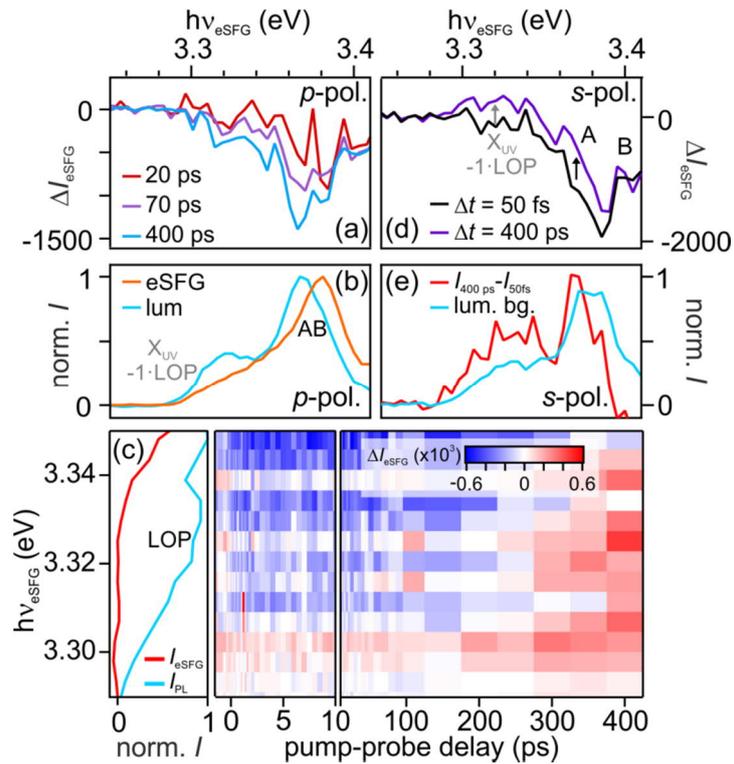

**Figure 5** (a) Change of eSFG intensity for *p*-polarization configuration on long ps timescales due to final state blocking by exciton ground state formation on a 100 ps timescale. The spectra resemble the static eSFG spectrum shown in (b) (orange), albeit with a different sign, and *not* the pump-induced static PL spectrum (light blue), which exhibits the LOP replica. (c) Photoinduced intensity change detected in the *s*-polarization configuration in false colors and as a function of pump-probe time delay at the energy of the LOP replica. Static PL and eSFG spectra are shown for comparison (left). (d) Lineouts from (c) at 50 fs (black) and 400 ps (purple). (e) The difference of the spectra (red) shown in (d) resembles the static PL spectrum (light blue), consistent with exciton ground state formation and stimulated PL by eSFG photons.

Figure 5c shows a zoom-in on the LOP replica of the dataset presented in Figure 4a after subtraction of the static eSFG spectrum and the PL background, which are plotted in the left panel for comparison. Clearly, the eSFG intensity strongly changes on long ps timescales. The spectral shape of the signal variation is depicted in Figure 5d. Right after photoexcitation ($\Delta t$ = 50 fs, black) we observe the negative intensity of the A/B resonance that is induced by the photodepletion of the intermediate defect states that was discussed in the framework of Figure 2d. At 400 ps (purple) the intensity at the energy of the A/B resonance increases and a positive signal appears at the position of the LOP replica. The photoinduced, delayed intensity change between 50 fs and 400 ps is plotted in Figure 5e (red). It shows a narrow peak centered at 3.37 eV and a broader one around 3.32 eV, the resonance energy of LOP-scattered excitons.

As the LOP-exciton resonance is not observed in the static eSFG spectra, its appearance in the time-resolved experiments must be correlated with the photoexcitation of ZnO by the pump pulse. Furthermore, the build-up of the LOP intensity occurs on the same timescale as the final state blocking of the A/B exciton-polariton resonance that we interpreted as the A/B exciton ground state formation. This unexpected observation can be rationalized by two different scenarios, which are both rate-limited by exciton ground state formation: Firstly, the photoinduced LOP intensity could be a result of sum frequency generation of two photons and involving the formation of a ground state A/B exciton coupling to a LOP, a highly non-linear process that seems very unlikely. Or, secondly, the eSFG photons at the A/B exciton-polariton resonance energy interact with ground state excitons and stimulate radiative recombination, leading to an enhanced PL intensity as soon as exciton ground state formation occurs. In the latter case, the photoinduced intensity distribution should reflect the spectral shape of the steady state PL.

Figure 5e compares the bare pump-induced luminescence spectrum (light blue) in the experiment to the photoinduced variation of the eSFG intensity (red). Both datasets agree well with each other. This suggests that the intensity variation is indeed related to an increase in photoluminescence rather than to a change of the eSFG signal. We propose that the temporal gating required for tr-PL is given by the eSFG photons, which trigger stimulated emission. Their presence induces the radiative recombination of excitons once they have reached their 1s ground level. As this is no longer a spontaneous effect, the exciton ground state formation is resolved in time.

Based on both observations, the final state blocking observed in tr-eSFG of the A/B and C resonances and the build-up of stimulated luminescence through eSFG photons, we conclude that exciton ground state formation occurs in the ZnO single crystal on a few 100 ps timescale, in accordance with previous time-resolved THz spectroscopy experiments that reported exciton formation within 200 ps.[34] It should be noted that this observation is not conflicting with the 200 fs timescale of surface exciton formation published in Ref. [51] and measured using the same ZnO sample. While the 200 fs time constant in our previous work is associated to the initial step of electron-hole pair formation, the population of a bound state in the hydrogen-like potential of the hole, the exciton *ground state* formation relates to the population of the 1s state, which is a prerequisite for efficient light emission.

## 3. Discussion and Conclusion

Figure 6 summarizes the main findings of the previous sections, which span a time range from 100 fs to 25 µs. Photoexcitation with sub-100 fs laser pulses at photon energies above the band gap energy of ZnO leads, on the one hand, to the depletion of electronic states in the defect band, forming positively charged defects. On the other hand, electrons from the topmost valence bands (A-C) are promoted to the CB, creating quasi-free carriers that relax to the band extrema. In the case of "hot" electrons in the CB, these dynamics can be as fast as 20 fs due to strong coupling to the LOP.[30,51] Thereby the carriers relax and the CB minimum is reached within 1 ps.[52] By comparison of the tr-EST data acquired using the defect-rich single crystal sample and the defect-poor MBE film, we find that these quasi-free electrons at the CB minimum exhibit defect-dependent lifetimes between 4 and 260 ps. In the case of the single crystal, the fast ($\leq 10$ ps) relaxation happens due to the formation of defect excitons $DX_{VIS}$ and defect-bound excitons $DX_{UV}$, while the slow decay (260 ps) in the MBE sample is likely to also contain contributions of free exciton formation, consistent with previous time-resolved THz spectroscopy experiments.[34] This delayed exciton formation is probably connected to the timescale of phonon scattering into the series of bound excitonic states (inset) and, in the case of the defect-related excitons, to the defect density, which is rate-limiting this process.

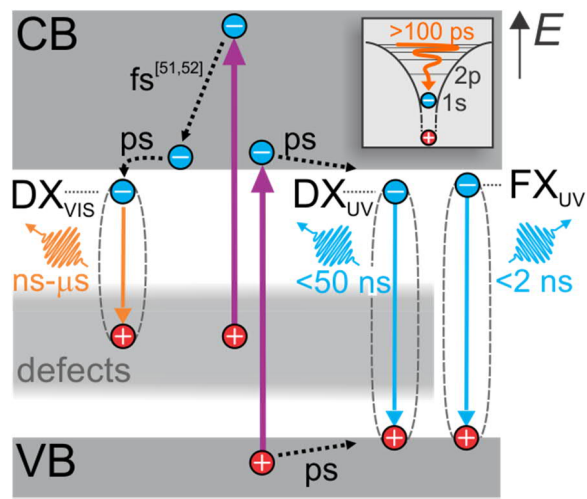

**Figure 6** Overview of the observed dynamics, ranging across more than eight orders of magnitude in time. Photoexcitation depletes normally occupied defect states forming $D^+$ with a µs lifetime and generates carriers in the CB and VB. Defect excitons ($DX_{VIS}$) and defect-bound excitons ($DX_{UV}$) are formed on ps timescales that depend on the defect density of the sample and decay within ns to µs.

Only if the defect density is sufficiently low and, thus, $DX_{UV}$ build-up sufficiently slow, free exciton ground state formation (on hundreds of ps timescales) and emission become competitive.

Our tr-PL measurements show that the free excitons decay radiatively within few ns, while the emission of defect-bound excitons $DX_{UV}$ continues up to 50 ns. The defect exciton $DX_{VIS}$ emission shows significantly longer dynamics that extend into the µs regime, which is connected to the long lifetime of the photodepleted defect states that we detect even after 25 µs. In addition, using tr-eSFG, we also observe decay times of holes at the top valence bands of few ps, which may be related to hole-polaron formation, which was discovered recently.[53] Lastly, we find that the exciton ground state formation occurs on a timescale larger than 100 ps.

These results illustrate the complexity of the non-equilibrium dynamics of charge carriers and excitons in photoexcited ZnO and highlight the crucial role of defects in this highly debated material. Defect states influence the optical properties of ZnO mainly in two ways: Firstly, the visible emission characteristics are a direct consequence of the positively charged defects that are formed abruptly upon photoexcitation of the sample. Secondly, depending on the defect density of the sample, the formation of defect-bound excitons competes with or even suppresses free exciton dynamics. If the defect density is sufficiently high, electron-hole pairs are trapped near defect sites, because this is energetically more favorable.

This two-fold impact of defects on the non-equilibrium properties of ZnO provides several interesting considerations for future technological applications of this recalcitrant material, depending on the desired functionality. Firstly, the initial photodepletion of defects and, thus, the ZnO emission in the visible spectrum depends on the defect density and the incident photon flux. Due to the energy level alignment also photon energies in the visible range are sufficient to generate this emission. The contribution of defect-bound excitons with emission characteristics near the intrinsic exciton resonances of ZnO depends critically on the defect density, too. However, beyond the pure correlation of the number of possible trapping sites (defects) to the number of defect-bound excitons, the formation dynamics actually *compete* with free exciton dynamics. This leads to significantly faster $FX_{UV}$ emission decay than the $DX_{UV}$ recombination. Subject to which application is envisioned, either fast light emission at the intrinsic resonance energy of ZnO may be desired or longer photoluminescence times, which can be tuned by the defect density in the material.

The defect density-dependent balance of $DX_{UV}$ as well as $FX_{UV}$ formation and emission dynamics is most likely also the origin of the divergence of previously published time constants of various processes in ZnO.[27-33] The Mott density, for example, is the critical charge carrier density at which the screening of the Coulomb interaction through free carriers suppresses exciton formation. As the defect density determines the lifetime of free carriers, a larger defect density, i.e. shorter free carrier lifetime, will lead to a much weaker reduction of the UV emission lines compared to a low defect density, albeit dominated by $DX_{UV}$ emission. Depending on the determination of the actual carrier densities in the experiment, huge divergences of deduced Mott density and Bohr exciton radius may result. For instance, defect-rich samples may appear to have a higher Mott density than defect-poor ZnO.

In addition, also the pump photon energy significantly affects the balance of visible and UV light emission: For pump photon energies below the ZnO band gap, photoexcitation across the gap requires multiphoton absorption. If the defect band were not present, the same absorbed amount of energy would lead to the same number of electron-hole pairs in the sample. However, as absorption of visible light from the defect band to the CB of ZnO is a linear process, it is significantly more likely than the non-linear excitation of carriers across the gap using visible photons. This is particularly the case for the commonly used pump photon energy of 1.55 eV (800 nm) that requires the absorption of three photons to excite across the ZnO band gap. In such experiments, the same amount of absorbed energy must lead to photodepletion of defects that is orders of magnitude larger than for UV photoexcitation. Differently put, the generation of the same number of electron-hole pairs in CB and VB requires orders of magnitude higher excitation fluences for near-IR or visible excitation photon energies than for UV excitation and causes an abundance of positively charged defects as well as quasi-free electrons in the CB with unknown consequences on the carrier and exciton dynamics.

In conclusion, this work provides new insights into the competition of intrinsic and defect-related exciton and charge carrier dynamics in ZnO, ranging over more than eight orders of magnitude in time. By combining three complementary time-resolved optical spectroscopic techniques, we are able to disentangle all relevant relaxation pathways in this challenging material. As a result, we are able to identify the main origin of previous conflicting findings: The defect density-dependent dynamics of the free carrier decay generating $DX_{VIS}$ and $DX_{UV}$ competing with free exciton formation. We propose to exploit our discoveries by manipulating the defect density in ZnO. Finally, it should be noted that the observed competition of the intrinsic charge carrier relaxation, exciton formation and decay dynamics with defect-induced effects is most probably not restricted to ZnO. Other TCOs and also compound semiconductors, in particular with wide band gaps like GaN or InGaN often expose comparably prominent defect bands and likely exhibit very similar non-equilibrium properties.

**Acknowledgements**

We thank Sylke Blumstengel and Prof. Dr. F. Henneberger for inspiring discussions as well as for contributing the static PL spectra shown in Figure 1a and 1b. This project was partially funded by the Deutsche Forschungsgemeinschaft (DFG, German Research Foundation) - Projektnummer 182087777 - SFB 951 and the European Commission through the FP7-NMP grant No. 280879-2 CRONOS.

**Appendix A: Experimental details**

The experiments happened under controlled temperature and pressure conditions of 100 K and $1\times10^{-7}$ mbar in an optical cryostat (Oxford Instruments, Optistat CF-V for EST, eSFG, and tr-PL; Microstat for static, high resolution PL spectroscopy). The ZnO single crystal is a hydrothermally grown crystal (MaTeck) cut along the non-polar ($10\bar{1}0$) surface and has a history of multiple *ex situ* sputtering and annealing cycles. It is compared to a 125 nm film grown by MBE on the ($10\bar{1}0$) surface of an unprepared single crystal (CrysTec) at a substrate temperature of 330 °C with further annealing to 700 °C. The charge carrier and exciton dynamics were investigated using complementary time-resolved optical techniques: PL spectroscopy, EST and eSFG.

Tr-EST and tr-eSFG were performed in a pump-probe scheme, where a first ultrashort laser pulse (pump) drives the system out of equilibrium and then, after a variable time delay, a second laser pulse (probe) interacts with the sample and monitors the pump-induced changes of the material properties. We used a Ti:Sa regeneratively amplified fs laser system (Coherent RegA) providing 1.55 eV photons in pulses of 45 fs duration. For tr-EST, the RegA was set to a repetition rate of 200 kHz (5 µs inter-pulse time difference) while for tr-eSFG it was decreased to 40 kHz (25 µs inter-pulse time difference). Note that, at these high repetition rates, all laser pulses hit one and the same spot on the sample. In contrast to pump-probe experiments in solution sample heating is not a problem, as heat is efficiently transported away from the spot in a bulk sample that is cooled by liquid nitrogen (100 K). Moreover, the repeated measurement enables the observation of long-lived states at negative pump-probe time delays where the probe precedes the pump on the ultrafast timescale, but effectively monitors the sample properties after 5 and 25 µs, respectively. For tr-EST and tr-eSFG, we used an optical parametric amplifier (OPA) providing output in the visible range and driven by 50% of the RegA power to generate the harmonic of the UV pump photon energy, which was achieved by frequency-doubling in a BBO crystal. For tr-EST, ZnO was photoexcited with $h\nu_{pump} = 3.8$ eV and an incident excitation fluence of 150 µJ cm$^{-2}$. In tr-eSFG, the pump photon energy was 4.1 eV, in order to shift the frequency of the stray light of the pump beam out of the eSFG photon energy range. The incident excitation fluence was kept below 23 µJ cm$^{-2}$.

The remaining RegA power was used to generate the probe pulses. In the case of tr-EST, the visible transmission was probed using a white light continuum in the 1.8 to 2.7 eV energy range, generated by focusing 10 % (≈1 µJ) of the RegA output in a 3 mm sapphire plate. The white light continuum was compressed down to 20 fs pulse duration using a deformable mirror,[54] such that the time resolution of the experiment was limited by the 45 fs width of the pump pulse. The transmitted white light was filtered by color filters (10 nm bandwidth) in the range between 540 nm (2.3 eV) and 694 nm (1.78 eV) and detected using a photodiode and a lock-in amplifier.

The eSFG experiments take advantage of the second order non-linear optical effects arising in the non-centrosymmetric bulk of ZnO. The technique requires the simultaneous interaction of two probe laser pulses in the sample: the broadband compressed white light continuum was up-converted by the remaining fundamental of the RegA to the 3.32-4.03 eV energy range, in order to be resonant with electronic transitions close to the ZnO band edge. The spectral features were resolved with an energy resolution that was limited by the 66 meV bandwidth of the 1.55 eV pulse. To avoid any spurious dynamics due to multiphoton absorption, the 1.55 eV light was reduced to 1.83 mJcm$^{-2}$. The orientation of the ZnO crystal in the cryostat aligned the *c*-axis ([0001] direction) in parallel to the *s*-polarization of the white light. The polarization of the up-converting beam was either perpendicular (*s*) or in parallel (*p*) to the plane of incidence (see inset of Figure 1b). The generated eSFG intensity was dispersed in a spectrometer (Andor Shamrock 303i) and detected in an electron multiplied charge coupled device (EMCCD).[43]

As illustrated by the inset in Figure 1b, in both experiments, tr-EST and tr-eSFG, the white light was incident at an angle of 45° (purple). In tr-EST, the pump beam was almost collinear with the white light (40°), as was the upconverting beam (43-44°) of the eSFG. The pump beam in tr-eSFG was aligned with a vertical angle of 45° to the surface normal (green).

To generate PL, the pump beam ($h\nu_{pump}$ = 3.8 eV) was focused into the optical cryostat, with incident excitation densities of 120 µJ cm$^{-2}$ in the case of the UV emission, 120 µJ cm$^{-2}$ and 230 µJ cm$^{-2}$ in the case of the VIS emission of the single crystal sample and the MBE film, respectively. Comparison of the PL spectra as a function of excitation fluence did not show any redshift of the exciton resonance for fluences below 500 µJ/cm$^2$ (not shown) that would be a sign for a significant influence of many body effects. The emitted light was collected outside the optical cryostat using a lense ($f$ = 5 cm) and focused into a monochromator (Bentham). Time resolution was achieved by TCSPC, which was performed using a hybrid photomultiplier detector (PicoQuant). This detector was triggered by the signal of a fast photodiode with a rise time shorter than 50 ps, which detected the arrival time of scattered light of the excitation beam.

It should be noted that for all pump photon energies used in the experiments, the absorption coefficient amounts to ca. 2·10$^5$ cm$^{-1}$.[55] The optical absorption depth is, thus, with 50 nm clearly shorter than the MBE film thickness (125 nm) and 92 % of the pump photons are absorbed in the epilayer.

**Appendix B: PL fit parameters**

The time-resolved measurements of the PL of both samples unveiled qualitative similarities and differences between the defect-rich and –poor sample, as discussed in detail in section 2.2. Although our work does not aim at a detailed analysis of the properties of specific defect-bound excitons, which would require elaborate characterization of different defect species and which is out of the scope of the present work, we provide the relevant fit functions and parameters in the following. The time-resolved PL data in Figure 2 was fitted using multiexponential fits of the form

$$I(t) = A_0 + \sum_i A_i \cdot \exp\left(-\frac{t}{\tau_i}\right) \quad (2)$$

convolved with a Gaussian that reproduces the main feature of the instrument response function (not shown). Table 1 summarizes the best fit parameters for the smallest number of exponential decays required to lead to satisfactory fit results. Note that the fits span four orders of magnitude in time. The quite high number of parameters in the case of the single crystal UV emission data can be reduced when using three stretched exponential functions to fit the data (not shown).

The relative intensity of the defect-related dynamics (cf. section 2.2) is determined by

$$I_{rel} = \frac{\sum A(\tau_i < 10 \text{ ns})}{\sum A(\tau_i > 10 \text{ ns})} \quad (3)$$

and shows that these dynamics have a $Q = I_{rel}^{SC} / I_{rel}^{MBE} = 30$ and 70 times higher relative intensity in the single crystal compared to the MBE-grown sample for UV and VIS emission, respectively

| | UV emission | | | | VIS emission | | | |
|---|---|---|---|---|---|---|---|---|
| | single crystal | | MBE film | | single crystal | | MBE film | |
| | $A_i$ (arb. u.) | $\tau_i$ (ps) | $A_i$ (arb. u.) | $\tau_i$ (ps) | $A_i$ (arb. u.) | $\tau_i$ (ps) | $A_i$ (arb. u.) | $\tau_i$ (ps) |
| i = 0 | 20(10) | - | 1.826 | - | 33.1(2) | - | 12.0(2) | - |
| 1 | 4034(7)e+3 | 15.71(4) | 5668e+3 | 144.8 | 3.62(3)e+3 | 51.5(6) | 58.0(2)e+3 | 16.9(1) |
| 2 | 171(2)e+3 | 237(3) | 1.057e+3 | 1.269e+3 | 824(7) | 710(7) | 8.68(5)e+3 | 169(2) |
| 3 | 108(1)e+3 | 1.21(3)e+3 | 112.9 | 21.77e+3 | 136(2) | 13.5(2)e+3 | 900(11) | 1.650(20)e+3 |
| 4 | 20(2)e+3 | 4.6(3)e+3 | - | - | 96.9(4) | 217(2)e+3 | 45.1(6) | 132(3)e+3 |
| 5 | 2.4(6)e+3 | 18(3)e+3 | - | - | - | - | - | - |
| 6 | 200(30) | 215(58)e+3 | - | - | - | - | - | - |
| $I_{rel}$ | 6e-4 | | 2e-5 | | 6e-2 | | 8.5e-4 | |
| Q | 30 | | | | 70 | | | |

**Table 1** Fit parameters of the traces shown in Figure 2. Errors smaller than 0.1 % were omitted for clarity.